\documentclass{optica-article}
\usepackage{appendix}
\journal{opticajournal}

\articletype{Research Article}

\usepackage{appendix}
\usepackage{lineno}
\usepackage{hyperref}
\usepackage{xcolor}
\begin{document}

\title{Arbitrary high-dimensional entanglement purification with the superposition of paths}

\author{Zi-Han Zheng\authormark{1}, Wen-Qiang Liu\authormark{1,*}, Chen-Ming Bai\authormark{1}, and Hai-Rui Wei\authormark{2}}

\address{\authormark{1}Department of Mathematics and Physics, Shijiazhuang Tiedao University, Shijiazhuang 050043, China\\
\authormark{2}School of Mathematics and Physics, University of Science and Technology Beijing, Beijing 100083, China}
\email{\authormark{*}wqliu@stdu.edu.cn}

\begin{abstract*}

High-dimensional quantum entanglement offers significant advantages over its two-dimensional counterparts in various quantum information processing tasks. We propose a scalable method for implementing an arbitrary high-dimensional entanglement purification protocol (HDEPP) using path superposition. By selecting appropriate single-qudit gates, qudit-flip errors and phase-flip errors can be purified. Counterintuitively, the HDEPP becomes more robust for higher-dimensional systems and exhibits enhanced purification fidelity as the dimension increases. Our HDEPP not only requires fewer operational resources but also in principle can achieve unity fidelity through multiple rounds of iteration. Furthermore, we develop feasible optical schemes to purify the frequency degree of freedom of photons suffering from qudit-flip and phase-flip errors using hyperentanglement. This work paves the way toward high-dimensional quantum communication and quantum networks.

\end{abstract*}

\section{Introduction} \label{sec1}

Quantum entanglement \cite{horodecki2009quantum} is an indispensable resource in quantum computation, quantum metrology, and quantum communication \cite{pan2001entanglement}. However, during practical preparation and distribution, entanglement is inevitably affected by noise arising from system-environment interactions and imperfect operations. This noise degrades the entanglement into low-quality mixed states, consequently impairing the performance of various quantum information processing tasks. Entanglement purification protocols (EPPs) \cite{yan2023advances} provide an effective strategy for distilling high-quality entangled states from noisy mixed states. In 1996, Bennett \emph{et al.} \cite{bennett1996purification} first proposed an EPP based on the controlled-NOT (CNOT) gate. Since then, a variety of EPPs have been theoretically proposed and experimentally demonstrated \cite{bennett1996purification,duan2000entanglement, duan2001long,dur2003multiparticle,sheng2008efficient, sheng2010one,zhou2020multi,krastanov2019optimized, hu2021long,wang2025single,zhou2020purification,zhou2025observation, PhysRevA.85.062326, PhysRevLett.110.260503, PhysRevA.104.012419, PhysRevA.105.062418, Yan2025quantum, PhysRevLett.127.040502, PhysRevA.109.042423}. These include gate-based EPPs \cite{bennett1996purification, duan2000entanglement}, linear-optical EPPs \cite{duan2001long, hu2021long}, cross-Kerr nonlinearity EPPs \cite{sheng2008efficient}, measurement-based EPP \cite{PhysRevA.85.062326, PhysRevLett.110.260503, PhysRevA.104.012419, PhysRevA.105.062418, Yan2025quantum}, entanglement-assisted EPP \cite{PhysRevLett.127.040502, PhysRevA.109.042423}, and hyperentanglement-based EPPs \cite{sheng2010one,hu2021long,wang2025single}. In addition, multi-partite EPPs \cite{dur2003multiparticle}, optimized EPPs \cite{krastanov2019optimized}, multi-copy EPPs \cite{zhou2020multi}, and residual entanglement in EPP \cite{zhou2020purification,zhou2025observation} have also been proposed, further enriching the toolbox for practical quantum information processing.

High-dimensional qudit systems with $d$-ary ($d > 2$) digits have been widely used to encode and process quantum information \cite{wang2020qudits} and show several important advantages over conventional two-dimensional qubit systems. For instance, they can significantly increase information capacity \cite{dixon2012quantum}, enhance resilience to environmental noise \cite{ecker2019overcoming}, improve the security of quantum communication \cite{cerf2002security}, boost the efficiency and accuracy of various quantum tasks \cite{campbell2014enhanced}, simplify quantum gate operations \cite{liu2020low}, and strengthen violations of Bell inequalities \cite{dada2011experimental}. In recent years, high-dimensional entanglement purification protocols (HDEPPs) have also been proposed \cite{horodecki1999reduction, vollbrecht2003efficient, cheong2007entanglement, miguel2018efficient, fang2025experimental, li2026double}. In 1999, Horodecki \emph{et al.} \cite{horodecki1999reduction} first extended the EPP to two-body high-dimensional systems. In 2007, Cheong \emph{et al.} \cite{cheong2007entanglement} presented an explicit EPP for high-dimensional Greenberger-Horne-Zeilinger states. In 2018, Miguel-Ramiro and D\"{u}r \cite{miguel2018efficient} proposed an HDEPP based on two-qudit generalized XOR gate operations. Subsequently, in 2025, Fang \emph{et al.} \cite{fang2025experimental} experimentally realized single-copy high-dimensional entanglement distillation. In 2026, Li \emph{et al.} \cite{li2026double} proposed a double-selection HDEPP using three copies of noisy entangled states and some generalized qudit operations.
However, most existing EPPs are still focused on two-dimensional qubit systems, and the currently available HDEPPs rely on multiple high-dimensional two-qudit gates, making their specific physical implementations challenging.

Unlike classical particles restricted to a single path, quantum particles can coherently traverse multiple paths simultaneously via coherent control. Such superpositions manifest as the superposition of spatial trajectories  \cite{chiribella2019quantum} and the superposition of causal order \cite{chiribella2013quantum}. Recently, the superposition of causal order has been shown to offer advantages in enhancing channel capacity \cite{wu2025general, deng2025generalized}, reducing quantum computational complexity \cite{liu2024experimentally, liu2025quantum, liu2026practical}, improving measurement precision \cite{yin2023experimental}, and enabling unambiguous discrimination of entangled states \cite{zhao2025unambiguous}. The superposition of paths \cite{chiribella2019quantum}, realized by coherently superposing two or more spatial trajectories, offers general communication advantages, including increased transmission capacity \cite{abbott2020communication}, improved teleportation \cite{Sayan2026path}, and the generation of noise-robust entanglement \cite{pellitteri2026entanglement}. More recently, a superposed EPP \cite{MiguelRamiro2025improving} and a distillation protocol employing higher-order superpositions of causal order \cite{kechrimparis2025probabilistic} were proposed to enhance purification efficiency. In contrast, the physical framework of spatial path superposition is well established and considerably easier to realize experimentally. Nevertheless, the study and potential advantages of spatial path superposition in the context of HDEPPs have not yet been explored.

In this paper, we first present a three-dimensional EPP for purifying qutrit-flip and phase-flip errors using the superposition of paths. By selecting appropriate single-qudit operations, our approach is then extended to realize an arbitrary HDEPP. Compared with previous HDEPP \cite{miguel2018efficient}, our protocol requires the same number of gate operations for purifying qudit-flip errors, but reduces four quantum Fourier-transform operations for phase-flip errors, thereby enhancing experimental feasibility.
Furthermore, leveraging hyperentanglement, we design two feasible optical schemes to purify the frequency degree of freedom (DOF) of photons suffering from qutrit-flip and phase-flip errors, respectively. These schemes can be extended to arbitrary high-dimensional systems. Finally, we analyze the success condition of our protocol and find that it is particularly suitable for higher-dimensional systems. Notably, as the dimension grows, the scheme achieves higher fidelity. Moreover, through iterative purification, the fidelity can in principle be improved to unity.

\section{EPP for two-qutrit Bell states} \label{Sec2}

\subsection{Qutrit-based EPP for correcting the qutrit-flip error}  \label{Sec2.1}

Suppose that a two-qutrit Bell state $|\Psi^{(3)}_{0,0}\rangle$ suffers from qutrit-flip errors during transmission to Alice and Bob. After transmission, the initial distributed state becomes a mixed state
\begin{eqnarray}            \label{eq1}
\rho_{AB}^{(3)}= F_0^{(3)}|\Psi_{0,0}^{(3)}\rangle_{AB}\langle\Psi_{0,0}^{(3)}|+F_1^{(3)}|\Psi_{0,1}^{(3)}\rangle_{AB}\langle\Psi_{0,1}^{(3)}|+F_2^{(3)}|\Psi_{0,2}^{(3)}\rangle_{AB}\langle\Psi_{0,2}^{(3)}|,
\end{eqnarray}
where the coefficients satisfy $F_0^{(3)}+F_1^{(3)}+F_2^{(3)}=1$. The subscripts $A$ and $B$ label the target qutrit states transmitted to Alice and Bob, respectively. The two-qutrit Bell states are given by
\begin{eqnarray}            \label{eq2}
\begin{split}
& |\Psi^{(3)}_{0,0}\rangle_{AB}=\frac{1}{\sqrt{3}}(|00\rangle+|11\rangle+|22\rangle)_{AB},\\
& |\Psi^{(3)}_{0,1}\rangle_{AB}=\frac{1}{\sqrt{3}}(|01\rangle+|12\rangle+|20\rangle)_{AB},\\
& |\Psi^{(3)}_{0,2}\rangle_{AB}=\frac{1}{\sqrt{3}}(|02\rangle+|10\rangle+|21\rangle)_{AB}.
\end{split}
\end{eqnarray}
\begin{figure}
\centering
\includegraphics[width=8.2 cm]{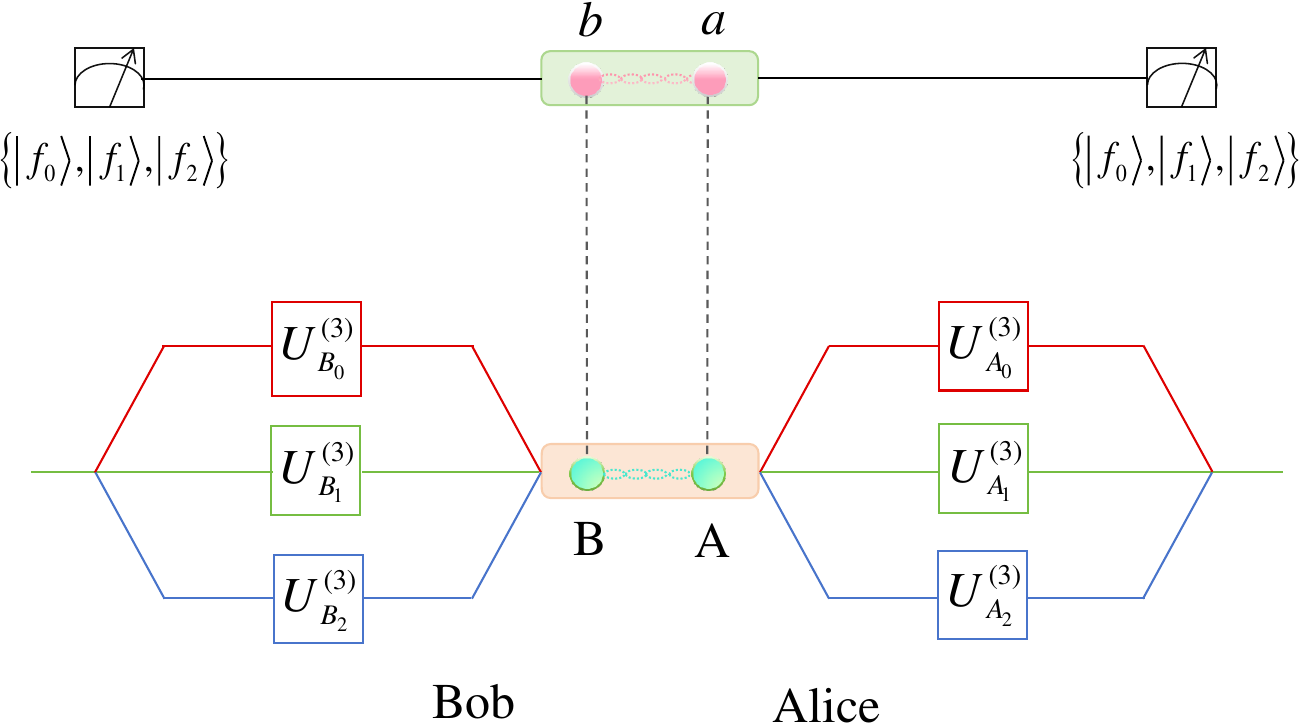}
\caption{Schematic diagram of the three-dimensional EPP for qutrit-flip or phase-flip errors using path superposition. An ancillary qutrit Bell state $|\Psi_{0,0}^{(3)}\rangle_{ab}$ is distributed to Alice and Bob and coherently controls the transmission paths of particles $A$ and $B$. Finally, qutrits $a$ and $b$ are measured in the Fourier basis and appropriate operations $U_{A_i}^{(3)}$ and $U_{B_i}^{(3)} (i=0, 1, 2)$ are chosen according to the type of error to be purified.}
\label{figure1}
\end{figure}

As illustrated in Fig. \ref{figure1}, we purify the state $|\Psi^{(3)}_{0,0}\rangle_{AB}$ from the mixed state in Eq. \eqref{eq1} by employing entanglement and coherent superposition of paths. Let us explain our protocol step by step.

\textbf{Step 1:} As shown in Fig. \ref{figure1}, an additional ancillary Bell state $|\Psi^{(3)}_{0,0}\rangle_{ab}$ and the target state $|\Psi^{(3)}_{0,0}\rangle_{AB}$ are distributed simultaneously to Alice and Bob. Suppose that after transmission the ancillary state undergoes a phase-flip error and the target state experiences a qutrit-flip error. Then the initial state $|\Psi^{(3)}_{0,0}\rangle_{ab}\otimes|\Psi^{(3)}_{0,0}\rangle_{AB}$ transforms into
\begin{eqnarray}             \label{eq3}
\rho^{(3)}=\rho_{ab}^{(3)}\otimes\rho_{AB}^{(3)}.
\end{eqnarray}
Here $\rho_{ab}^{(3)}$ is a mixed state after transmission, given by
\begin{eqnarray}             \label{eq4}
\rho_{ab}^{(3)}=\tilde{F}_0^{(3)}|\Psi_{0,0}^{(3)}\rangle_{ab}\langle\Psi_{0,0}^{(3)}|+\tilde{F}_1^{(3)}|\Psi_{1,0}^{(3)}\rangle_{ab}\langle\Psi_{1,0}^{(3)}|+\tilde{F}_2^{(3)}|\Psi_{2,0}^{(3)}\rangle_{ab}\langle\Psi_{2,0}^{(3)}|,
\end{eqnarray}
with the qutrit Bell states
\begin{eqnarray}            \label{eq5}
\begin{split}
&|\Psi^\text{(3)}_{1,0}\rangle_{ab}=\frac{1}{\sqrt{3}}(|00\rangle+e^{\frac{2\pi}{3} \texttt{i}}|11\rangle+e^{\frac{\pi}{3} \texttt{i}}|22\rangle)_{ab},\\
&|\Psi^\text{(3)}_{2,0}\rangle_{ab}=\frac{1}{\sqrt{3}}(|00\rangle+e^{\frac{\pi}{3} \texttt{i}}|11\rangle+e^{\frac{2\pi}{3} \texttt{i}}|22\rangle)_{ab}.
\end{split}
\end{eqnarray}
The subscripts $a$ and $b$ denote the ancillary states transmitted to Alice and Bob, respectively. Therefore, the initial state becomes a probabilistic mixture of nine states: $|\Psi_{i,j}^{(3)}\rangle'=|\Psi_{i,0}^{(3)}\rangle_{ab}\otimes|\Psi_{0,j}^{(3)}\rangle_{AB}$ with a probability of $\tilde{F}_i^{(3)}F_j^{(3)}$, where $i,j=0,1,2$.

\textbf{Step 2:} The controlled operations are applied to the system. Specifically, the qutrits $a$ and $b$ serve as the control qutrits and coherently control the qutrits $A$ and $B$, respectively. If qutrit $a$ is in the state $|k\rangle_a$, the single-qutrit operation $U_{A_k}^{(3)}$ is applied to qutrit $A$. Bob performs an analogous operation on qutrits $b$ and $B$. These controlled operations are given by

%
\begin{eqnarray}             \label{eq6}
S_{aA}^{(3)}=\sum_{k=0}^2|k\rangle_a\langle k| \otimes U_{A_k}^{(3)} ,\quad S_{bB}^{(3)}=\sum_{k=0}^2|k\rangle_b\langle k| \otimes U_{B_k}^{(3)}.
\end{eqnarray}
%

%
%
%
%

%
%
Hence, the overall controlled operation is $S^{(3)}=S_{aA}^{(3)}\otimes S_{bB}^{(3)}.$ Applying $S^{(3)}$ to the nine states $|\Psi_{i,j}^{(3)}\rangle'$ yields the transformed states $|\tilde{\Psi}_{i,j}^{(3)}\rangle' = S^{(3)}|\Psi_{i,j}^{(3)}\rangle'$, which are given by
\begin{eqnarray}
\begin{split}             \label{eq7}
|\tilde{\Psi}_{0,0}^{(3)}\rangle'=
&\frac{1}{\sqrt{3}}\big(|0\rangle_a\otimes|0\rangle_b\otimes U_{A_0}^{(3)}\otimes U_{B_0}^{(3)}+|1\rangle_a\otimes|1\rangle_b\otimes U_{A_1}^{(3)}\\&\otimes U_{B_1}^{(3)}+|2\rangle_a\otimes|2\rangle_b\otimes U_{A_2}^{(3)}\otimes U_{B_2}^{(3)}\big)|\Psi_{0,0}^{(3)}\rangle_{AB},
\end{split}
\end{eqnarray}
\begin{eqnarray}
\begin{split}             \label{eq8}
|\tilde{\Psi}_{0,1}^{(3)}\rangle'=
&\frac{1}{\sqrt{3}}\big(|0\rangle_a\otimes|0\rangle_b\otimes U_{A_0}^{(3)}\otimes U_{B_0}^{(3)}+|1\rangle_a\otimes|1\rangle_b\otimes U_{A_1}^{(3)}\\&\otimes U_{B_1}^{(3)}+|2\rangle_a\otimes|2\rangle_b\otimes U_{A_2}^{(3)}\otimes U_{B_2}^{(3)}\big)|\Psi_{0,1}^{(3)}\rangle_{AB},
\end{split}
\end{eqnarray}
\begin{eqnarray}
\begin{split}             \label{eq9}
|\tilde{\Psi}_{0,2}^{(3)}\rangle'=
&\frac{1}{\sqrt{3}}\big(|0\rangle_a\otimes|0\rangle_b\otimes U_{A_0}^{(3)}\otimes U_{B_0}^{(3)}+|1\rangle_a\otimes|1\rangle_b\otimes U_{A_1}^{(3)}\\&\otimes U_{B_1}^{(3)}+|2\rangle_a\otimes|2\rangle_b\otimes U_{A_2}^{(3)}\otimes U_{B_2}^{(3)}\big)|\Psi_{0,2}^{(3)}\rangle_{AB},
\end{split}
\end{eqnarray}
\begin{eqnarray}
\begin{split}             \label{eq10}
|\tilde{\Psi}_{1,0}^{(3)}\rangle'=
&\frac{1}{\sqrt{3}}\big(|0\rangle_a\otimes|0\rangle_b\otimes U_{A_0}^{(3)}\otimes U_{B_0}^{(3)}+e^{\frac{2\pi}{3} \texttt{i}}|1\rangle_a\otimes|1\rangle_b\otimes U_{A_1}^{(3)}\\&\otimes U_{B_1}^{(3)}+e^{\frac{\pi}{3} \texttt{i}}|2\rangle_a\otimes|2\rangle_b\otimes U_{A_2}^{(3)}\otimes U_{B_2}^{(3)}\big)|\Psi_{0,0}^{(3)}\rangle_{AB},
\end{split}
\end{eqnarray}
\begin{eqnarray}
\begin{split}             \label{eq11}
|\tilde{\Psi}_{1,1}^{(3)}\rangle'=
&\frac{1}{\sqrt{3}}\big(|0\rangle_a\otimes|0\rangle_b\otimes U_{A_0}^{(3)}\otimes U_{B_0}^{(3)}+e^{\frac{2\pi}{3} \texttt{i}}|1\rangle_a\otimes|1\rangle_b\otimes U_{A_1}^{(3)}\\&\otimes U_{B_1}^{(3)}+e^{\frac{\pi}{3} \texttt{i}}|2\rangle_a\otimes|2\rangle_b\otimes U_{A_2}^{(3)}\otimes U_{B_2}^{(3)}\big)|\Psi_{0,1}^{(3)}\rangle_{AB},
\end{split}
\end{eqnarray}
\begin{eqnarray}
\begin{split}             \label{eq12}
|\tilde{\Psi}_{1,2}^{(3)}\rangle'=
&\frac{1}{\sqrt{3}}\big(|0\rangle_a\otimes|0\rangle_b\otimes U_{A_0}^{(3)}\otimes U_{B_0}^{(3)}+e^{\frac{2\pi}{3} \texttt{i}}|1\rangle_a\otimes|1\rangle_b\otimes U_{A_1}^{(3)}\\&\otimes U_{B_1}^{(3)}+e^{\frac{\pi}{3} \texttt{i}}|2\rangle_a\otimes|2\rangle_b\otimes U_{A_2}^{(3)}\otimes U_{B_2}^{(3)}\big)|\Psi_{0,2}^{(3)}\rangle_{AB},
\end{split}
\end{eqnarray}
\begin{eqnarray}
\begin{split}             \label{eq13}
|\tilde{\Psi}_{2,0}^{(3)}\rangle'=
&\frac{1}{\sqrt{3}}\big(|0\rangle_a\otimes|0\rangle_b\otimes U_{A_0}^{(3)}\otimes U_{B_0}^{(3)}+e^{\frac{\pi}{3} \texttt{i}}|1\rangle_a\otimes|1\rangle_b\otimes U_{A_1}^{(3)}\\
&\otimes U_{B_1}^{(3)}+e^{\frac{2\pi}{3} \texttt{i}}|2\rangle_a\otimes|2\rangle_b\otimes U_{A_2}^{(3)}\otimes U_{B_2}^{(3)}\big)|\Psi_{0,0}^{(3)}\rangle_{AB},
\end{split}
\end{eqnarray}
\begin{eqnarray}
\begin{split}             \label{eq14}
|\tilde{\Psi}_{2,1}^{(3)}\rangle'=
&\frac{1}{\sqrt{3}}\big(|0\rangle_a\otimes|0\rangle_b\otimes U_{A_0}^{(3)}\otimes U_{B_0}^{(3)}+e^{\frac{\pi}{3} \texttt{i}}|1\rangle_a\otimes|1\rangle_b\otimes U_{A_1}^{(3)}\\&\otimes U_{B_1}^{(3)}+e^{\frac{2\pi}{3} \texttt{i}}|2\rangle_a\otimes|2\rangle_b\otimes U_{A_2}^{(3)}\otimes U_{B_2}^{(3)}\big)|\Psi_{0,1}^{(3)}\rangle_{AB},
\end{split}
\end{eqnarray}
\begin{eqnarray}
\begin{split}             \label{eq15}
|\tilde{\Psi}_{2,2}^{(3)}\rangle'=
&\frac{1}{\sqrt{3}}\big(|0\rangle_a\otimes|0\rangle_b\otimes U_{A_0}^{(3)}\otimes U_{B_0}^{(3)}+e^{\frac{\pi}{3} \texttt{i}}|1\rangle_a\otimes|1\rangle_b\otimes U_{A_1}^{(3)}\\&\otimes U_{B_1}^{(3)}+e^{\frac{2\pi}{3} \texttt{i}}|2\rangle_a\otimes|2\rangle_b\otimes U_{A_2}^{(3)}\otimes U_{B_2}^{(3)}\big)|\Psi_{0,2}^{(3)}\rangle_{AB}.
\end{split}
\end{eqnarray}

\textbf{Step 3:} Finally, Alice and Bob each measure their ancillary qutrits $a$ and $b$ in the Fourier basis $\{|f_0\rangle, |f_1\rangle, |f_2\rangle\}$. The Fourier basis states are expressed in the computational basis as
\begin{eqnarray}
\begin{split}             \label{eq16}
&|f_0\rangle=\frac{1}{\sqrt{3}}(|0\rangle+|1\rangle+|2\rangle),\\
&|f_1\rangle=\frac{1}{\sqrt{3}}(|0\rangle+e^{\frac{2\pi}{3} \texttt{i}}|1\rangle+e^{\frac{\pi}{3} \texttt{i}}|2\rangle),\\
&|f_2\rangle=\frac{1}{\sqrt{3}}(|0\rangle+e^{\frac{\pi}{3} \texttt{i}}|1\rangle+e^{\frac{2\pi}{3} \texttt{i}}|2\rangle).
\end{split}
\end{eqnarray}
To purify the qutrit-flip error, we choose the single-qutrit operations $U_{A_i}^{(3)}$ and $U_{B_i}^{(3)}$ to be the three-dimensional generalized Pauli $Z_3$ operators, specifically
\begin{eqnarray}
\begin{split}             \label{eq17}
&U_{A_1}^{(3)}=U_{B_1}^{(3)}=I=\sum_{\ell=0}^2|\ell\rangle\langle\ell|,\\
&U_{A_2}^{(3)}=U_{B_0}^{(3)} = Z_3^1=\sum_{\ell=0}^2e^{\frac{2\pi\ell}{3} \texttt{i}}|\ell\rangle\langle\ell|,\\
&U_{A_0}^{(3)}=U_{B_2}^{(3)}=Z_3^2=\sum_{\ell=0}^2e^{\frac{\pi\ell}{3} \texttt{i}}|\ell\rangle\langle\ell|. \\
\end{split}
\end{eqnarray}

Substituting Eq.~\eqref{eq17} into Eqs.~\eqref{eq7}–\eqref{eq15} and applying the Fourier basis definition in Eq.~\eqref{eq16}, the states in Eqs.~\eqref{eq7}–\eqref{eq15} simplify to
\begin{eqnarray}             \label{eq18}
|\tilde{\Psi}_{0,0}^{(3)}\rangle'=\frac{1}{\sqrt{3}}(|f_0\rangle_a|f_0\rangle_b+|f_1\rangle_a|f_2\rangle_b+|f_2\rangle_a|f_1\rangle_b)\otimes|\Psi_{0,0}^{(3)}\rangle_{AB},
\end{eqnarray}
\begin{eqnarray}             \label{eq19}
|\tilde{\Psi}_{0,1}^{(3)}\rangle'=\frac{1}{\sqrt{3}}(|f_0\rangle_a|f_2\rangle_b+|f_1\rangle_a|f_1\rangle_b+|f_2\rangle_a|f_0\rangle_b)\otimes|\Psi_{0,1}^{(3)}\rangle_{AB},
\end{eqnarray}
\begin{eqnarray}             \label{eq20}
|\tilde{\Psi}_{0,2}^{(3)}\rangle'=\frac{1}{\sqrt{3}}(|f_0\rangle_a|f_1\rangle_b+|f_1\rangle_a|f_0\rangle_b+|f_2\rangle_a|f_2\rangle_b)\otimes|\Psi_{0,2}^{(3)}\rangle_{AB},
\end{eqnarray}
\begin{eqnarray}             \label{eq21}
|\tilde{\Psi}_{1,0}^{(3)}\rangle'=\frac{1}{\sqrt{3}}(|f_0\rangle_a|f_1\rangle_b+|f_1\rangle_a|f_0\rangle_b+|f_2\rangle_a|f_2\rangle_b)\otimes|\Psi_{0,0}^{(3)}\rangle_{AB},
\end{eqnarray}
\begin{eqnarray}             \label{eq22}
|\tilde{\Psi}_{1,1}^{(3)}\rangle'=\frac{1}{\sqrt{3}}(|f_0\rangle_a|f_0\rangle_b+|f_1\rangle_a|f_2\rangle_b+|f_2\rangle_a|f_1\rangle_b)\otimes|\Psi_{0,1}^{(3)}\rangle_{AB},
\end{eqnarray}
\begin{eqnarray}             \label{eq23}
|\tilde{\Psi}_{1,2}^{(3)}\rangle'=\frac{1}{\sqrt{3}}(|f_0\rangle_a|f_2\rangle_b+|f_1\rangle_a|f_1\rangle_b+|f_2\rangle_a|f_0\rangle_b)\otimes|\Psi_{0,2}^{(3)}\rangle_{AB},
\end{eqnarray}
\begin{eqnarray}             \label{eq24}
|\tilde{\Psi}_{2,0}^{(3)}\rangle'=\frac{1}{\sqrt{3}}(|f_0\rangle_a|f_2\rangle_b+|f_1\rangle_a|f_1\rangle_b+|f_2\rangle_a|f_0\rangle_b)\otimes|\Psi_{0,0}^{(3)}\rangle_{AB},
\end{eqnarray}
\begin{eqnarray}             \label{eq25}
|\tilde{\Psi}_{2,1}^{(3)}\rangle'=\frac{1}{\sqrt{3}}(|f_0\rangle_a|f_1\rangle_b+|f_1\rangle_a|f_0\rangle_b+|f_2\rangle_a|f_2\rangle_b)\otimes|\Psi_{0,1}^{(3)}\rangle_{AB},
\end{eqnarray}
\begin{eqnarray}             \label{eq26}
|\tilde{\Psi}_{2,2}^{(3)}\rangle'=\frac{1}{\sqrt{3}}(|f_0\rangle_a|f_0\rangle_b+|f_1\rangle_a|f_2\rangle_b+|f_2\rangle_a|f_1\rangle_b)\otimes|\Psi_{0,2}^{(3)}\rangle_{AB}.
\end{eqnarray}

When the measurement results are $|f_0\rangle_a|f_0\rangle_b$, $|f_1\rangle_a|f_2\rangle_b$, or $|f_2\rangle_a |f_1\rangle_b$, then the resulting new mixed state is
\begin{eqnarray}             \label{eq27}
\rho_1^{(3)}=L_{0,0}^{(3)}|\Psi^{(3)}_{0,0}\rangle_{AB}\langle\Psi^{(3)}_{0,0}|+L_{0,1}^{(3)}|\Psi^{(3)}_{0,1}\rangle_{AB}\langle\Psi^{(3)}_{0,1}|+L_{0,2}^{(3)}|\Psi^{(3)}_{0,2}\rangle_{AB}\langle\Psi^{(3)}_{0,2}|,
\end{eqnarray}
where
\begin{eqnarray}            \label{eq28}
L_{0,0}^{(3)}=\frac{\tilde{F}_0^{(3)}F_0^{(3)}}{\sum_{i=0}^2\tilde{F}_i^{(3)}F_i^{(3)}},\quad L_{0,1}^{(3)}=\frac{\tilde{F}_1^{(3)}F_1^{(3)}}{\sum_{i=0}^2\tilde{F}_i^{(3)}F_i^{(3)}},\quad L_{0,2}^{(3)}=1-L_{0,0}^{(3)}-L_{0,1}^{(3)}.
\end{eqnarray}

\subsection{Qutrit-based EPP for correcting the phase-flip error}  \label{Sec2.2}

Suppose that a two-qutrit Bell state $|\Psi_{0,0}^{(3)}\rangle$ suffers from phase-flip error during transmission to Alice and Bob. After transmission, the initial distributed state becomes a mixed state
\begin{eqnarray}             \label{eq29}
\bar{\rho}_{AB}^{(3)}=\bar{F}_0^{(3)}|\Psi_{0,0}^{(3)}\rangle_{AB}\langle\Psi_{0,0}^{(3)}|+\bar{F}_1^{(3)}|\Psi_{1,0}^{(3)}\rangle_{AB}\langle\Psi_{1,0}^{(3)}|+\bar{F}_2^{(3)}|\Psi_{2,0}^{(3)}\rangle_{AB}\langle\Psi_{2,0}^{(3)}|,
\end{eqnarray}
where the coefficients satisfy $\bar{F}_0^{(3)}+\bar{F}_1^{(3)}+\bar{F}_2^{(3)}=1$. As shown in Fig. \ref{figure1}, to purify the state $|\Psi_{0,0}^{(3)}\rangle_{AB}$, we introduce an ancillary distributed state $|\Psi_{0,0}^{(3)}\rangle_{ab}$. Suppose that after transmission the ancillary state also undergoes a phase-flip error. The initial state $|\Psi_{0,0}^{(3)}\rangle_{ab}\otimes|\Psi_{0,0}^{(3)}\rangle_{AB}$ then transforms into
\begin{eqnarray}             \label{eq30}
\bar{\rho}^{(3)}=\rho_{ab}^{(3)}\otimes \bar{\rho}_{AB}^{(3)}.
\end{eqnarray}
Therefore, the initial state becomes a probabilistic mixture of nine states: $|\bar{\Psi}_{i,j}^{(3)}\rangle=|\Psi_{i,0}^{(3)}\rangle_{ab}\otimes|\Psi_{j,0}^{(3)}\rangle_{AB}$ with a probability of $\tilde{F}_i^{(3)}\bar{F}_j^{(3)}$.

Similar to the purification of the qutrit-flip error in step 2, we apply $S^{(3)}$ operation to the nine states, i.e., $|\bar{\Psi}_{i,j}^{(3)}\rangle'= S^{(3)}|\bar{\Psi}_{i,j}^{(3)}\rangle$. To purify the phase-flip error, the operations $U_{A_i}^{(3)}$ and $U_{B_i}^{(3)}$ are chosen as the three-dimensional generalized Pauli $X_3$ operations, specifically,
\begin{eqnarray}
\begin{split}             \label{eq31}
&U_{A_0}^{(3)}=U_{B_0}^{(3)}=I=\sum_{\ell=0}^2|\ell\rangle\langle\ell|, \\
&U_{A_1}^{(3)}=U_{B_1}^{(3)}=X_3^1=\sum_{\ell=0}^2|\ell\oplus1\rangle\langle\ell|,\\
&U_{A_2}^{(3)}=U_{B_2}^{(3)}= X_3^2=\sum_{\ell=0}^2|\ell\oplus2\rangle\langle\ell|. \\
\end{split}
\end{eqnarray}
where $\ell\oplus1\equiv (\ell+1) \bmod d,\ell\oplus2\equiv (\ell+2) \bmod d$ and $d=3$. Substituting Eq.~\eqref{eq31} into $|\bar{\Psi}_{i,j}^{(3)}\rangle'$, states $|\bar{\Psi}_{i,j}^{(3)}\rangle'$ simplify to
\begin{eqnarray}             \label{eq32}
|\bar{\Psi}_{0,0}^{(3)}\rangle'=\frac{1}{\sqrt{3}}(|f_0\rangle_a|f_0\rangle_b+|f_1\rangle_a|f_2\rangle_b+|f_2\rangle_a|f_1\rangle_b)\otimes|\Psi_{0,0}^{(3)}\rangle_{AB},
\end{eqnarray}
\begin{eqnarray}             \label{eq33}
|\bar{\Psi}_{0,1}^{(3)}\rangle'=\frac{1}{\sqrt{3}}(|f_0\rangle_a|f_2\rangle_b+|f_1\rangle_a|f_1\rangle_b+|f_2\rangle_a|f_0\rangle_b)\otimes|\Psi_{1,0}^{(3)}\rangle_{AB},
\end{eqnarray}
\begin{eqnarray}             \label{eq34}
|\bar{\Psi}_{0,2}^{(3)}\rangle'=\frac{1}{\sqrt{3}}(|f_0\rangle_a|f_1\rangle_b+|f_1\rangle_a|f_0\rangle_b+|f_2\rangle_a|f_2\rangle_b)\otimes|\Psi_{2,0}^{(3)}\rangle_{AB},
\end{eqnarray}
\begin{eqnarray}             \label{eq35}
|\bar{\Psi}_{1,0}^{(3)}\rangle'=\frac{1}{\sqrt{3}}(|f_0\rangle_a|f_1\rangle_b+|f_1\rangle_a|f_0\rangle_b+|f_2\rangle_a|f_2\rangle_b)\otimes|\Psi_{0,0}^{(3)}\rangle_{AB},
\end{eqnarray}
\begin{eqnarray}             \label{eq36}
|\bar{\Psi}_{1,1}^{(3)}\rangle'=\frac{1}{\sqrt{3}}(|f_0\rangle_a|f_0\rangle_b+|f_1\rangle_a|f_2\rangle_b+|f_2\rangle_a|f_1\rangle_b)\otimes|\Psi_{1,0}^{(3)}\rangle_{AB},
\end{eqnarray}
\begin{eqnarray}             \label{eq37}
|\bar{\Psi}_{1,2}^{(3)}\rangle'=\frac{1}{\sqrt{3}}(|f_0\rangle_a|f_2\rangle_b+|f_1\rangle_a|f_1\rangle_b+|f_2\rangle_a|f_0\rangle_b)\otimes|\Psi_{2,0}^{(3)}\rangle_{AB},
\end{eqnarray}
\begin{eqnarray}             \label{eq38}
|\bar{\Psi}_{2,0}^{(3)}\rangle'=\frac{1}{\sqrt{3}}(|f_0\rangle_a|f_2\rangle_b+|f_1\rangle_a|f_1\rangle_b+|f_2\rangle_a|f_0\rangle_b)\otimes|\Psi_{0,0}^{(3)}\rangle_{AB},
\end{eqnarray}
\begin{eqnarray}             \label{eq39}
|\bar{\Psi}_{2,1}^{(3)}\rangle'=\frac{1}{\sqrt{3}}(|f_0\rangle_a|f_1\rangle_b+|f_1\rangle_a|f_0\rangle_b+|f_2\rangle_a|f_2\rangle_b)\otimes|\Psi_{1,0}^{(3)}\rangle_{AB},
\end{eqnarray}
\begin{eqnarray}             \label{eq40}
|\bar{\Psi}_{2,2}^{(3)}\rangle'=\frac{1}{\sqrt{3}}(|f_0\rangle_a|f_0\rangle_b+|f_1\rangle_a|f_2\rangle_b+|f_2\rangle_a|f_1\rangle_b)\otimes|\Psi_{2,0}^{(3)}\rangle_{AB}.
\end{eqnarray}

Finally, Alice and Bob measure the ancillary qutrits $a$ and $b$ in the Fourier basis, respectively. When the measurement results are $|f_0\rangle_a|f_0\rangle_b$, $ |f_1\rangle_a|f_2\rangle_b$, or $|f_2\rangle_a|f_1\rangle_b$, one obtains a new mixed state
\begin{eqnarray}             \label{eq41}
\bar{\rho}_1^{(3)}=\bar{L}_{0,0}^{(3)}|\Psi^{(3)}_{0,0}\rangle_{AB}\langle\Psi^{(3)}_{0,0}|+\bar{L}_{0,1}^{(3)}|\Psi^{(3)}_{1,0}\rangle_{AB}\langle\Psi^{(3)}_{1,0}|+\bar{L}_{0,2}^{(3)}|\Psi^{(3)}_{2,0}\rangle_{AB}\langle\Psi^{(3)}_{2,0}|,
\end{eqnarray}
%
where
\begin{eqnarray}
\begin{split}             \label{eq42}
\bar{L}_{0,0}^{(3)}=\frac{\tilde{F}_0^{(3)}\bar{F}_0^{(3)}}{\sum_{i=0}^2\tilde{F}_i^{(3)}\bar F_i^{(3)}},
\quad\bar{L}_{0,1}^{(3)}=\frac{\tilde{F}_1^{(3)}\bar{F}_1^{(3)}}{\sum_{i=0}^2\tilde{F}_i^{(3)}\bar F_i^{(3)}},
\quad \bar{L}_{0,2}^{(3)}=1-\bar{L}_{0,0}^{(3)}-\bar{L}_{0,1}^{(3)}.
\end{split}
\end{eqnarray}

\section{EPP for two-qudit $d$-dimensional Bell states}  \label{Sec3}

\subsection{Qudit-based EPP for correcting the qudit-flip error}  \label{Sec3.1}

A $d$-dimensional Bell state with two parameters $n, m\in \{0, 1, \ldots, d-1\}$ is defined as
\begin{eqnarray}            \label{eq43}
|\Psi_{n,m}^{(d)}\rangle =\frac{1}{\sqrt{d}}\sum_{k=0}^{d-1}e^{2\pi\texttt{i}nk/d}|k\rangle_A|(k+m)\bmod d\rangle_B.
\end{eqnarray}

Suppose that a $d$-dimensional Bell state $|\Psi_{0,0}^{(d)}\rangle$ suffers from qudit-flip errors during transmission to Alice and Bob. After transmission, the initial distributed state becomes a mixed state
\begin{eqnarray}            \label{eq44}
\rho_{AB}^{(d)}=\sum_{k=0}^{d-1}F_k^{(d)}|\Psi_{0,k}^{(d)}\rangle_{AB}\langle\Psi_{0,k}^{(d)}|,
\end{eqnarray}
where the coefficients satisfy $\sum_{k=0}^{d-1}F_k^{(d)}=1$. Our method can be generalized to the arbitrary $d$-dimensional  EPP. Let us explain below the EPP step by step.
\begin{figure} 
\begin{center}
\includegraphics[width=8.2 cm]{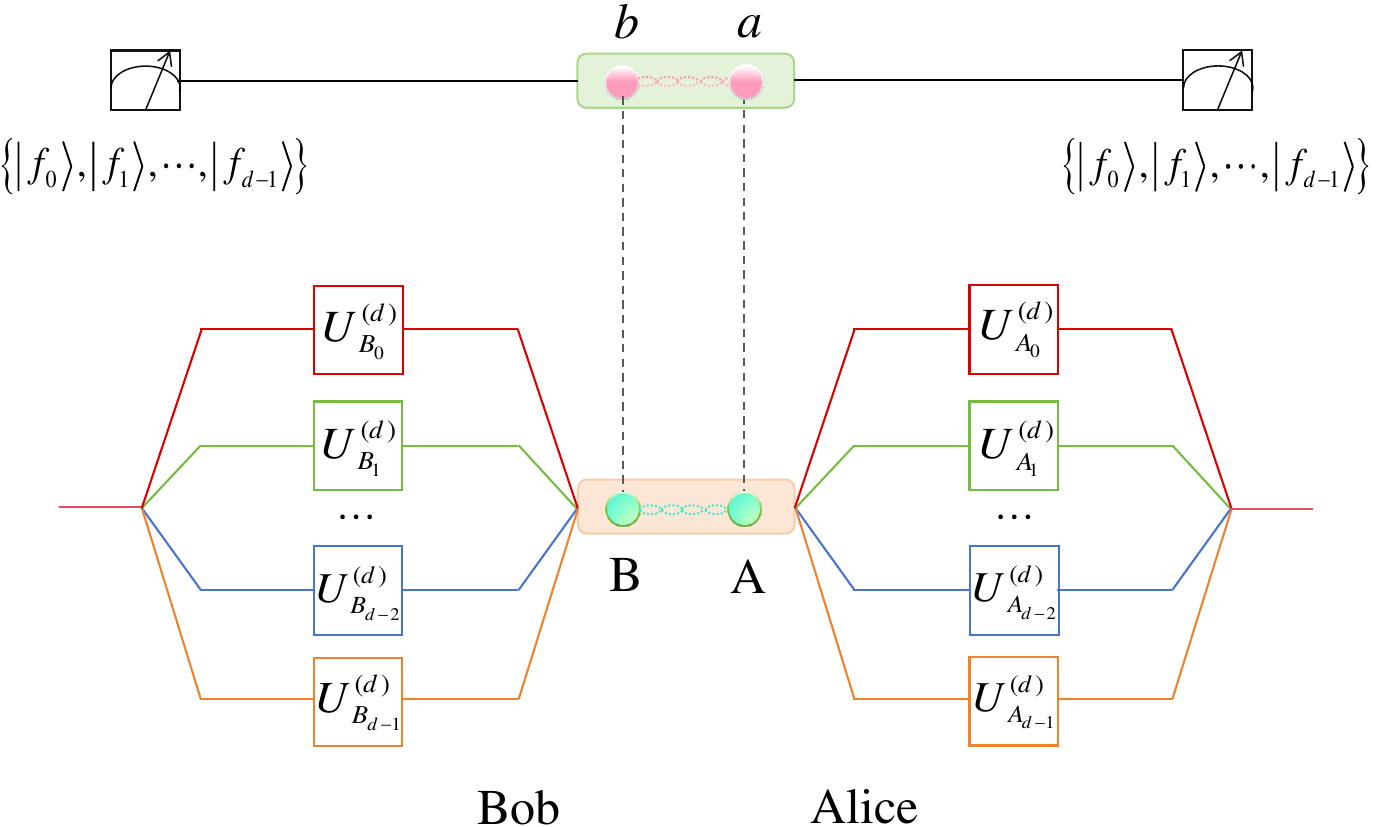}
\caption{Schematic diagram of the $d$-dimensional EPP for qutrit-flip or phase-flip errors using path superposition. Appropriate single-qudit operations $U_{A_k}^{(d)}$ and $U_{B_k}^{(d)}$ are chosen according to the type of error to be purified.}     \label{figure2}
\end{center}
\end{figure}

\textbf{Step 1:} As shown in Fig. \ref{figure2}, an additional ancillary Bell state $|\Psi_{0,0}^{(d)}\rangle_{ab}$ and the target state $|\Psi_{0,0}^{(d)}\rangle_{AB}$ are distributed simultaneously to Alice and Bob, where the ancillary state undergoes a phase-flip error and the target state experiences a qudit-flip error. So the initial state $|\Psi_{0,0}^{(d)}\rangle_{ab}\otimes|\Psi_{0,0}^{(d)}\rangle_{AB}$ transforms into
\begin{eqnarray}            \label{eq45}
\rho^{(d)}=\rho_{ab}^{(d)}\otimes\rho_{AB}^{(d)}.
\end{eqnarray}
Here $\rho_{ab}^{(d)}$ is a mixed state after transmission, given by
\begin{eqnarray}            \label{eq46}
\rho_{ab}^{(d)}=\sum_{l=0}^{d-1}\tilde{F}_l^{(d)}|\Psi_{l,0}^{(d)}\rangle_{ab}\langle\Psi_{l,0}^{(d)}|.
\end{eqnarray}
Therefore, the initial state becomes a probabilistic mixture of $d^2$ states. These states are $|\Psi_{l,k}^{(d)}\rangle'=|\Psi_{l,0}^{(d)}\rangle_{ab}\otimes|\Psi_{0,k}^{(d)}\rangle_{AB}$ with a probability of $\tilde{F}_l^{(d)}F_k^{(d)}$ ($l, k=0, 1, \ldots, d-1$).

\textbf{Step 2:} Two path-superposition controlled operations are applied to the system, given by
\begin{eqnarray}            \label{eq47}
S_{aA}^{(d)}=\sum_{r=0}^{d-1}|r\rangle_a\langle r|\otimes U_{A_r}^{(d)},\quad S_{bB}^{(d)}=\sum_{s=0}^{d-1}|s\rangle_b\langle s|\otimes U_{B_s}^{(d)}.
\end{eqnarray}
After the whole controlled operation $S^{(d)}=S_{aA}^{(d)}\otimes S_{bB}^{(d)}$ is applied to the $d^2$ mixed states, respectively. $|\tilde{\Psi}_{l,k}^{(d)}\rangle'=S^{(d)}|\Psi_{l,k}^{(d)}\rangle'$, where $\omega_{d}=e^{\frac{2\pi}{d}\texttt{i}}$ and one obtains
\begin{eqnarray}            \label{eq48}
|\tilde{\Psi}_{l,k}^{(d)}\rangle'=\frac{1}{\sqrt{d}}\sum_{r=0}^{d-1}
\big(\omega^{r\cdot l}_{d}|r\rangle_a\otimes|r\rangle_b\otimes U_{A_r}^{(d)}\otimes U_{B_r}^{(d)}\big) |\Psi_{0,k}^{(d)}\rangle_{AB}.
\end{eqnarray}
%

\textbf{Step 3:} Finally, Alice and Bob measure their ancillary qudits $a$ and $b$ in the $d$-dimensional Fourier basis $\{|f_0\rangle, |f_1\rangle, \ldots, |f_{d-1}\rangle\}$, respectively. The Fourier basis is related to the computational basis by the Fourier transform matrix $\mathcal{F}_d$:
\begin{eqnarray}             \label{eq49}
(|f_0\rangle,|f_1\rangle,\cdots,|f_{d-1}\rangle)^{\mathrm{T}}
=\mathcal{F}_d(|0\rangle,|1\rangle,\cdots,|d-1\rangle)^{\mathrm{T}},
\end{eqnarray}
where $\mathcal{F}_d$ is the $d\times d$ matrix
\begin{eqnarray}
\begin{split}             \label{eq50}
\mathcal{F}_d=\frac{1}{\sqrt{d}}\sum_{\mathscr{p}=0}^{d-1}\sum_{\mathscr{q}=0}^{d-1}\omega_d^{\mathscr{p}\mathscr{q}}|\mathscr{p}\rangle\langle\mathscr{q}|.\\
\end{split}
\end{eqnarray}

To purify the qudit-flip error, we select single-qudit gates $U_{A_k}^{(d)}$ and $U_{B_k}^{(d)}$ as the $d$-dimensional generalized Pauli $Z_d$ operation and its integer powers, i.e.,
\begin{eqnarray}            \label{eq51}
U_{A_k}^{(d)}=Z_d^{k-1}, \quad U_{B_k}^{(d)}=Z_d^{1-k}.
\end{eqnarray}
Here the $d$-dimensional generalized Pauli $Z_d$ operation is $Z_d=\sum_{\ell=0}^{d-1}\omega^{\ell}|\ell\rangle\langle\ell|.$
%

Substituting Eqs.~\eqref{eq52} and \eqref{eq54} into Eq.~\eqref{eq51}, the final state becomes
\begin{eqnarray}            \label{eq52}
|\tilde{\Psi}_{l,k}^{(d)}\rangle'=&\frac{1}{\sqrt{d}}\sum_{r=0}^{d-1}|f_r\rangle_a|f_{(d-r-k+l)\bmod d}\rangle_b\otimes|\Psi_{0,k}^{(d)}\rangle_{AB}.
\end{eqnarray}
If the measurement results are $|f_0\rangle_a|f_{(d-r)\bmod d}\rangle_b$, $|f_1\rangle_a|f_{(d-1-r)\bmod d}\rangle_b, \cdots, |f_{d-1}\rangle_a|f_{(1-r)\bmod d}\rangle_b$ $(r=l=0, 1, \ldots, d-1)$, one obtains the new mixed state
\begin{eqnarray}            \label{eq53}
\rho_{l+1}^{(d)}=\sum_{k=0}^{d-1}L_{l,k}^{(d)}|\Psi_{0,k}^{(d)}\rangle_{AB}\langle\Psi_{0,k}^{(d)}|,
\end{eqnarray}
where
\begin{eqnarray}            \label{eq54}
L_{l,k}^{(d)}=\frac{\tilde{F}_{(k-l)\bmod d}^{(d)}F_k^{(d)}}{\sum_{r=0}^{d-1}\tilde{F}_{(r-l)\bmod d}^{(d)}F_r^{(d)}}.
\end{eqnarray}

\subsection{Qudit-based EPP for correcting the phase-flip error}  \label{Sec3.2}

Suppose that a $d$-dimensional Bell state $|\Psi_{0,0}^{(d)}\rangle$ suffers from phase-flip error during transmission to Alice and Bob. After transmission, the initial distributed state becomes a mixed state
\begin{eqnarray}            \label{eq55}
\bar{\rho}_{AB}^{(d)}=\sum_{k=0}^{d-1}\bar{F}_k^{(d)}|\Psi_{k,0}^{(d)}\rangle_{AB}\langle\Psi_{k,0}^{(d)}|,
\end{eqnarray}
where the coefficients satisfy $\sum_{k=0}^{d-1}\bar{F}_k^{(d)}=1$. To purify state $|\Psi_{0,0}^{(d)}\rangle_{AB}$, an ancillary state $|\Psi_{0,0}^{(d)}\rangle_{ab}$ is distributed to Alice and Bob that also experiences a phase-flip error. The initial state $|\Psi_{0,0}^{(d)}\rangle_{ab}\otimes|\Psi_{0,0}^{(d)}\rangle_{AB}$ becomes
\begin{eqnarray}            \label{eq56}
\bar{\rho}^{(d)}=\rho_{ab}^{(d)}\otimes\bar{\rho}_{AB}^{(d)}.
\end{eqnarray}
Equation \eqref{eq68} is a probabilistic mixture of $d^2$ states. These states are $|\bar{\Psi}_{l,k}^{(d)}\rangle=|\Psi_{l,0}^{(d)}\rangle_{ab}\otimes|\Psi_{k,0}^{(d)}\rangle_{AB}$ with a probability of $\tilde{F}_l^{(d)}\bar{F}_k^{(d)}$.

Analogous to the purification of $d$-dimensional qudit-flip errors, the path-superposition controlled operation is applied to the system to evolve the state, yielding $|\bar{\Psi}_{l,k}^{(d)}\rangle'=S^{(d)}|\bar{\Psi}_{l,k}^{(d)}\rangle$.

Finally, the ancillary qudits are measured in the Fourier basis. To purify the phase-flip errors, the single-qudit gates $U_{A_k}^{(d)}$ and $U_{B_k}^{(d)}$ are chosen as $d$-dimensional generalized Pauli $X_d$ operation and its $k$-th powers $(k=0, 1, \ldots, d-1)$, i.e.,
\begin{eqnarray}            \label{eq57}
U_{A_k}^{(d)}=U_{B_k}^{(d)}=X_d^{k}.
\end{eqnarray}
The $d$-dimensional generalized Pauli $X_d$ operation is given by $X_d=\sum_{\ell=0}^{d-1}|\ell\oplus1\rangle\langle\ell|.$
%
%
So the state $|\bar{\Psi}_{l,k}^{(d)}\rangle'$ is explicitly given by
\begin{eqnarray}            \label{eq58}
|\bar{\Psi}_{l,k}^{(d)}\rangle'=\frac{1}{\sqrt{d}}\sum_{r=0}^{d-1}|f_r\rangle_a|f_{(d-r-k+l)\bmod d}\rangle_b\otimes|\Psi_{k,0}^{(d)}\rangle_{AB}.
\end{eqnarray}
If the measurement results are $|f_0\rangle_a|f_{(d-r)\bmod d}\rangle_b$, $|f_1\rangle_a|f_{(d-1-r)\bmod d}\rangle_b, \cdots, |f_{d-1}\rangle_a|f_{(1-r)\bmod d}\rangle_b$ $(r=l=0, 1, \ldots, d-1)$, one obtains the new mixed state
\begin{eqnarray}            \label{eq59}
\bar{\rho}_{l+1}^{(d)}=\sum_{k=0}^{d-1}\bar{L}_{l,k}^{(d)}|\Psi_{k,0}^{(d)}\rangle_{AB}\langle\Psi_{k,0}^{(d)}|,
\end{eqnarray}
where
\begin{eqnarray}            \label{eq60}
\begin{split}
\bar{L}_{l,k}^{(d)}=\frac{\tilde{F}_{(k-l)\bmod d}^{(d)}\bar{F}_k^{(d)}}{\sum_{r=0}^{d-1}\tilde{F}_{(r-l)\bmod d}^{(d)} \bar{F}_r^{(d)}}.
\end{split}
\end{eqnarray}

\section{Optical implementation}  \label{Sec4}

\subsection{Optical architecture of EPP for correcting qutrit-flip error}  \label{Sec4.1}

\begin{figure} 
\begin{center}
\includegraphics[width=8.2 cm]{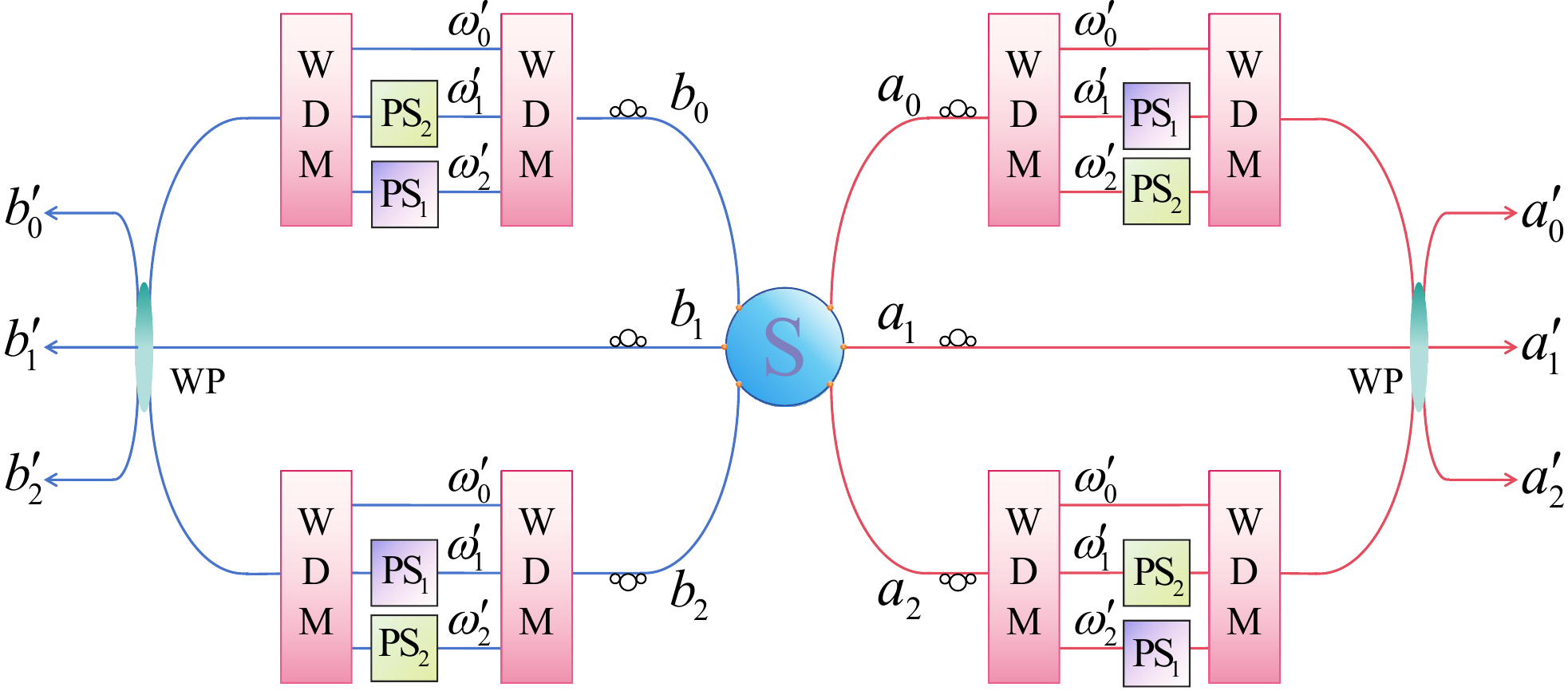}
\caption{Optical implementation diagram for purifying qutrit-flip errors on the frequency DOF of photons. The hyperentanglement source (S) generates a pair of spatial-frequency hyperentangled states. WDM denotes wavelength division multiplexing, which separates and combines photon frequencies. $\text{PS}_1$ and $\text{PS}_2$ are phase shifters that introduce the relative phases $e^{\frac{\pi}{3} \texttt{i}}$ and $e^{\frac{2\pi}{3} \texttt{i}}$, respectively. The wave plate (WP), acting as a tritter, performs the Fourier transform on the spatial-mode DOF.}    \label{figure3}
\end{center}
\end{figure}

To purify the qutrit-flip errors in frequency DOF, our optical EPP employs hyperentanglement encoded in spatial and frequency modes of photons. As shown in Fig. \ref{figure3}, the hyperentanglement source (S) first generates one pair of spatial-frequency hyperentangled states $|\Psi_{0,0}^{(3)}\rangle_s\otimes|\Psi_{0,0}^{(3)}\rangle_f$. Here the frequency entanglement $|\Psi_{0,0}^{(3)}\rangle_f$ is the target state to be distributed, while the spatial-mode entanglement $|\Psi_{0,0}^{(3)}\rangle_s$ serves as the ancillary state.

Since the spatial-mode DOF is immune to qutrit-flip errors, we assume that after transmission through the noisy channel, the spatial mode undergoes phase-flip errors, whereas the frequency mode suffers from qutrit-flip errors. Then the initial states $|\Psi_{00}^{(3)}\rangle_s\otimes|\Psi_{00}^{(3)}\rangle_f$ transform into
\begin{eqnarray}    \label{eq61}
\rho^{(3)}_{sf}=\rho_s^{(3)}\otimes\rho_f^{(3)}.
\end{eqnarray}
Here $\rho_s^{(3)}$ and $\rho_f^{(3)}$ are the noisy mixed states in the spatial and frequency DOFs, respectively,
\begin{eqnarray}    \label{eq62}
\rho_s^{(3)}=\tilde{F}_0^{(3)}|\Psi_{0,0}^{(3)}\rangle_s\langle\Psi_{0,0}^{(3)}|+\tilde{F}_1^{(3)}|\Psi_{1,0}^{(3)}\rangle_s\langle\Psi_{1,0}^{(3)}|+\tilde{F}_2^{(3)}|\Psi_{2,0}^{(3)}\rangle_s\langle\Psi_{2,0}^{(3)}|,
\end{eqnarray}
\begin{eqnarray}    \label{eq63}
\rho_f^{(3)}=F_0^{(3)}|\Psi_{0,0}^{(3)}\rangle_f\langle\Psi_{0,0}^{(3)}|+F_1^{(3)}|\Psi_{0,1}^{(3)}\rangle_f\langle\Psi_{0,1}^{(3)}|+F_2^{(3)}|\Psi_{0,2}^{(3)}\rangle_f\langle\Psi_{0,2}^{(3)}|.
\end{eqnarray}
The spatial and frequency qutrit Bell states are
\begin{eqnarray}
\begin{split}    \label{eq64}
&|\Psi_{0,0}^{(3)}\rangle_s=\frac{1}{\sqrt{3}}(|a_0\rangle|b_0\rangle+|a_1\rangle|b_1\rangle+|a_2\rangle|b_2\rangle),\\
&|\Psi_{1,0}^{(3)}\rangle_s=\frac{1}{\sqrt{3}}(|a_0\rangle|b_0\rangle+e^{\frac{2\pi}{3} \texttt{i}}|a_1\rangle|b_1\rangle+e^{\frac{\pi}{3} \texttt{i}}|a_2\rangle|b_2\rangle),\\
&|\Psi_{2,0}^{(3)}\rangle_s=\frac{1}{\sqrt{3}}(|a_0\rangle|b_0\rangle+e^{\frac{\pi}{3} \texttt{i}}|a_1\rangle|b_1\rangle+e^{\frac{2\pi}{3} \texttt{i}}|a_2\rangle|b_2\rangle).
\end{split}
\end{eqnarray}
\begin{eqnarray}
\begin{split}    \label{eq645}
&|\Psi_{0,0}^{(3)}\rangle_f=\frac{1}{\sqrt{3}}(|\omega_0'\rangle|\omega_0'\rangle+|\omega_1'\rangle|\omega_1'\rangle+|\omega_2'\rangle|\omega_2'\rangle),
\\&|\Psi_{0,1}^{(3)}\rangle_f=\frac{1}{\sqrt{3}}(|\omega_0'\rangle|\omega_1'\rangle+|\omega_1'\rangle|\omega_2'\rangle+|\omega_2'\rangle|\omega_0'\rangle),
\\&|\Psi_{0,2}^{(3)}\rangle_f=\frac{1}{\sqrt{3}}(|\omega_0'\rangle|\omega_2'\rangle+|\omega_1'\rangle|\omega_0'\rangle+|\omega_2'\rangle|\omega_1'\rangle).
\end{split}
\end{eqnarray}
Here the $a_i, b_i$ denote the spatial DOF of the photons, and $\omega_i'$ denotes their frequency DOF.

As shown in Fig.~\ref{figure3}, to implement the path-superposition controlled operation, the spatial DOF is used as the control qutrit, and the single-qutrit operations $U_{A_i}^{(3)}$ and $U_{B_i}^{(3)}$ (defined in Eq.~\eqref{eq17}) acting on the frequency DOF are applied in each spatial mode.  The $Z_3^1$ and $Z_3^2$ operations on the frequency DOF can be realized by a block consisting of two wavelength division multiplexers (WDMs) and two phase shifters, PS$_1$ and PS$_2$.
The first WDM separates different frequencies into distinct spatial modes, and PS$_1$ and PS$_2$ on the corresponding spatial mode pick up the relative phases $e^{\frac{\pi}{3} \texttt{i}}$ and $e^{\frac{2\pi}{3} \texttt{i}}$, respectively. The second WDM then combines the different frequencies back into the same spatial mode.

Finally, a wave plate (WP) acting
as a tritter (triport beam splitter) implements the Fourier transform on the spatial-mode DOF  \cite{PhysRevLett.73.58,kumar2023experimental} on Alice's side, i.e.,
\begin{eqnarray}
\begin{split}             \label{eq65}
&|a_0'\rangle=\frac{1}{\sqrt{3}}(|a_0\rangle+|a_1\rangle+|a_2\rangle),\\
&|a_1'\rangle=\frac{1}{\sqrt{3}}(|a_0\rangle+e^{\frac{2\pi}{3} \texttt{i}}|a_1\rangle+e^{\frac{\pi}{3} \texttt{i}}|a_2\rangle),\\
&|a_2'\rangle=\frac{1}{\sqrt{3}}(|a_0\rangle+e^{\frac{\pi}{3} \texttt{i}}|a_1\rangle+e^{\frac{2\pi}{3} \texttt{i}}|a_2\rangle).
\end{split}
\end{eqnarray}
Bob performs the same operations. By selecting the case where two photons in output modes $a_0'b_0'$, $a_1'b_2'$, or $a_2'b_1'$, the photonic state becomes a new mixed state
\begin{eqnarray}            \label{eq66}
\rho_f'^{(3)}=L_{0,0}^{(3)}|\Psi^{(3)}_{0,0}\rangle_f\langle\Psi^{(3)}_{0,0}|+L_{0,1}^{(3)}|\Psi^{(3)}_{0,1}\rangle_f\langle\Psi^{(3)}_{0,1}|+L_{0,2}^{(3)}|\Psi^{(3)}_{0,2}\rangle_f\langle\Psi^{(3)}_{0,2}|,
\end{eqnarray}
which corresponds to Eq. \eqref{eq27}. Similarly, by selecting the case where two photons in output modes $a_0'b_2'$, $a_1'b_1'$, or $a_2'b_0'$, and the case $a_0'b_1'$, $a_1'b_0'$, or $a_2'b_2'$, the resulting new mixed states in the frequency DOF like Eq. \eqref{eq27}.

\subsection{Optical architecture of EPP for correcting phase-flip error}  \label{Sec4.2}

\begin{figure} 
\begin{center}
\includegraphics[width=8.5 cm]{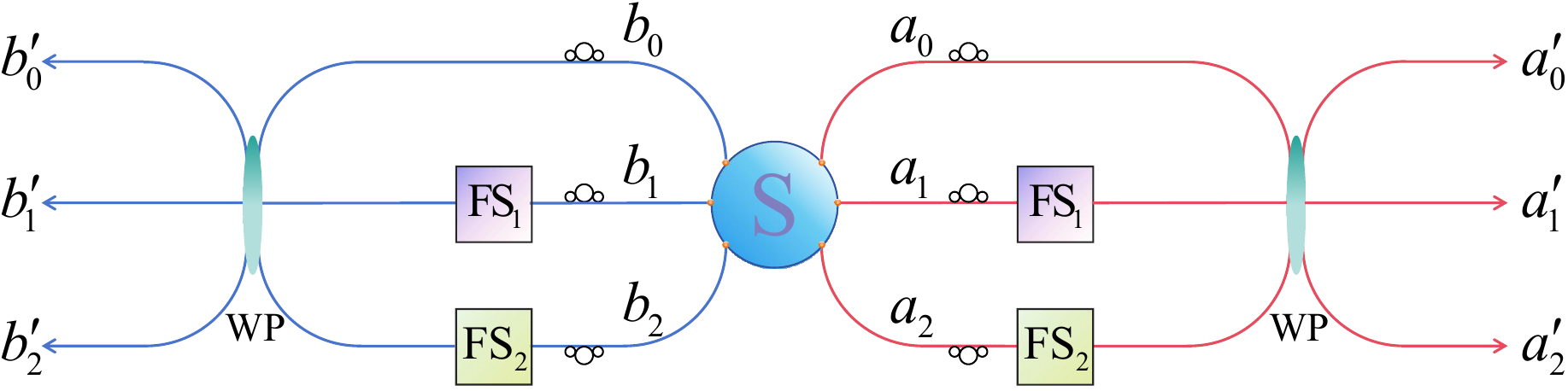}
\caption{Optical implementation diagram for purifying qutrit phase-flip errors on the frequency DOF of photons. Here $\text{FS}_1$ and $\text{FS}_2$ denote frequency shifters.}     \label{figure4}
\end{center}
\end{figure}

We consider a qutrit EPP for phase-flip errors on the frequency DOF. As shown in Fig. \ref{figure4}, the source (S) first generates a pair of spatial and frequency hyperentangled states $|\Psi_{0,0}^{(3)}\rangle_s\otimes|\Psi_{0,0}^{(3)}\rangle_f$. After the spatial-frequency DOFs are subject to phase-flip noise, and become
\begin{eqnarray}    \label{eq67}
\bar{\rho}^{(3)}_{sf}=\rho_s^{(3)}\otimes\bar{\rho}_f^{(3)}.
\end{eqnarray}
Here $\bar{\rho}_f^{(3)}$ is the mixed state on the frequency DOF with phase-flip noise, given by
\begin{eqnarray}    \label{eq68}
\bar{\rho}_f^{(3)}=\bar{F}_0^{(3)}|\Psi_{0,0}^{(3)}\rangle_f\langle\Psi_{0,0}^{(3)}|+\bar{F}_1^{(3)}|\Psi_{1,0}^{(3)}\rangle_f\langle\Psi_{1,0}^{(3)}|+\bar{F}_2^{(3)}|\Psi_{2,0}^{(3)}\rangle_f\langle\Psi_{2,0}^{(3)}|.
\end{eqnarray}
Here the qutrit Bell states on the frequency DOF are
\begin{eqnarray}
\begin{split}    \label{eq69}
&|\Psi_{1,0}^{(3)}\rangle_f=\frac{1}{\sqrt{3}}(|\omega_0'\rangle|\omega_0'\rangle+e^{\frac{2\pi}{3} \texttt{i}}|\omega_1'\rangle|\omega_1'\rangle+e^{\frac{\pi}{3} \texttt{i}}|\omega_2'\rangle|\omega_2'\rangle),
\\&|\Psi_{2,0}^{(3)}\rangle_f=\frac{1}{\sqrt{3}}(|\omega_0'\rangle|\omega_0'\rangle+e^{\frac{\pi}{3} \texttt{i}}|\omega_1'\rangle|\omega_1'\rangle+e^{\frac{2\pi}{3} \texttt{i}}|\omega_2'\rangle|\omega_2'\rangle).
\end{split}
\end{eqnarray}

To implement the path-superposition operation, the corresponding single-qutrit gates defined in Eq.~\eqref{eq31} are applied in each spatial mode. The $X_3^1$ and $X_3^2$ gates are implemented by the frequency shifters FS$_1$ and FS$_2$. Specifically, FS$_1$ changes the frequency as $\omega_0' \to \omega_1'$, $\omega_1' \to \omega_2'$, and $\omega_2' \to \omega_0'$, while FS$_2$ changes the frequency as $\omega_0' \to \omega_2'$, $\omega_1' \to \omega_0'$, and $\omega_2' \to \omega_1'$.

Finally, Alice and Bob both perform the Fourier transform on the spatial-mode DOF. By selecting the cases where the photons in output modes $a_0'b_0'$, $a_1'b_2'$, or $a_2'b_1'$, the resulting new mixed states in the frequency DOF correspond to Eq.~\eqref{eq41}.

\section{Discussion and conclusion}  \label{Sec5}

\begin{figure} [t]   
\centering
\includegraphics[width=8.2 cm]{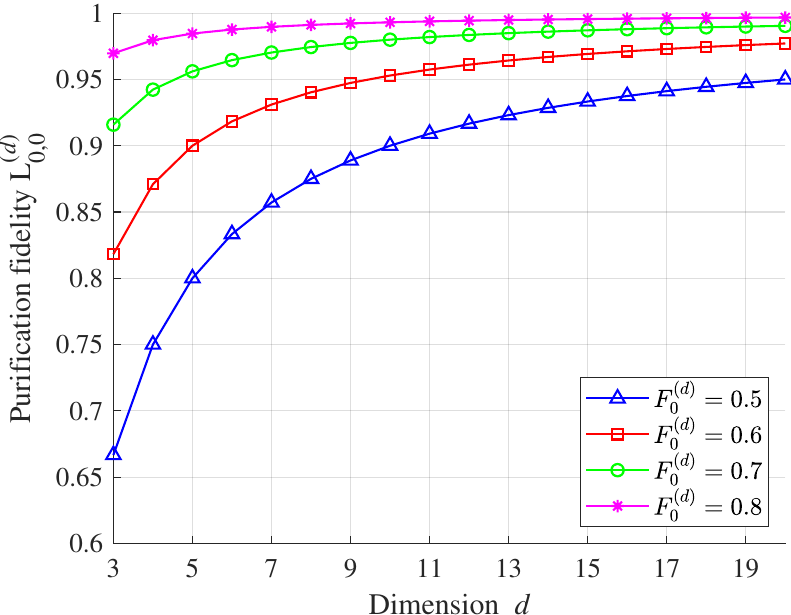}
\caption{Relationship between the purification fidelity $L_{0,0}^{(d)}$ and the dimension $d$ for different initial fidelities $F_{0}^{(d)}$, where $F_1^{(d)}=F_2^{(d)}=\cdots =F_{d-1}^{(d)}=\frac{1-F_0^{(d)}}{d-1}$ are taken.}     \label{FIG.5}
\end{figure}

The protocol succeeds if and only if the fidelity of the target state in the new mixed state is higher than that in the original noisy mixed state. For the measurement results $|f_0\rangle_a|f_0\rangle_b$, $|f_1\rangle_a|f_{d-1}\rangle_b$, $\ldots$, $|f_{d-1}\rangle_a|f_1\rangle_b$ and $F_r^{(d)}=\tilde{F}_r^{(d)}=\bar{F}_r^{(d)}$, the success condition of our $d$-level EPP for both qudit-flip and phase-flip errors is given by
\begin{eqnarray}   \label{eq70}
L_{0,0}^{(d)}=\bar{L}_{0,0}^{(d)}=\frac{(F_0^{(d)})^2}{\sum_{r=0}^{d-1}(F_r^{(d)})^2}>F_0^{(d)}.
\end{eqnarray}
Based on the Cauchy-Schwarz inequality, Eq.~\eqref{eq70} yields the success condition $F_0^{(d)}\in(\frac{1}{d},1]$ and the purification success probability is $\frac{1}{d}$.  Figure~\ref{FIG.5} presents the purification fidelity $L_{0,0}^{(d)}$ as a function of the dimension $d$, and indicates that higher-dimensional systems achieve higher fidelity. This behavior originates from the dilution of errors: as shown in Eq. \eqref{eq70}, as $d$ increases, the error probability per branch decreases, allowing the post-selection to suppress error components more effectively. For measurement outcomes other than $|f_0\rangle_a|f_0\rangle_b$, $|f_1\rangle_a|f_{d-1}\rangle_b$, $\ldots$, $|f_{d-1}\rangle_a|f_1\rangle_b$, the success condition becomes $F_0^{(d)}\in[0,\frac{1}{d})$ and purification success probability is $\frac{d-1}{d}$.

\begin{figure} [t] 
\centering
\includegraphics[width=8.2 cm]{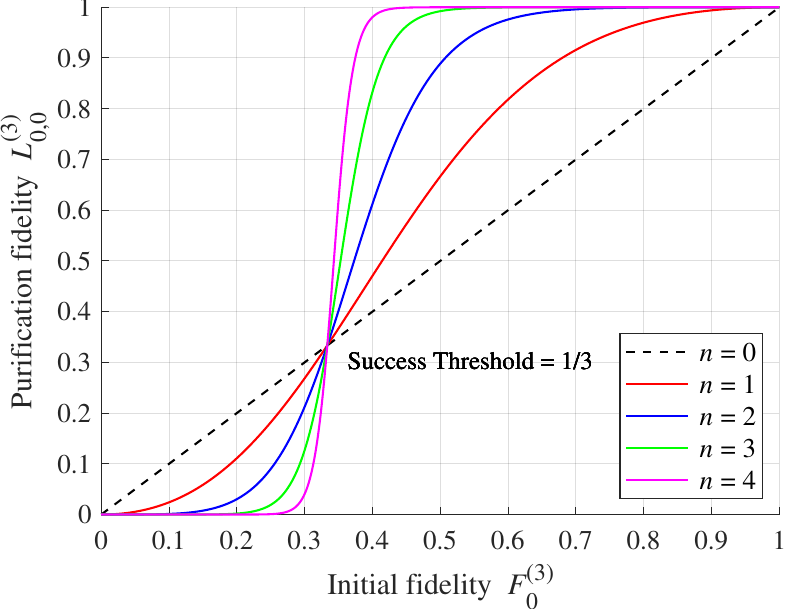}
\caption{Purification fidelity $L_{0,0}^{(3)}$ as a function of the initial fidelity $F_0^{(3)}$ for $d=3$, with different iteration numbers $n$.}     \label{FIG.6}
\end{figure}

In our protocol, the purification fidelity can be improved by iterating the EPP processes. After
$n$ rounds of iteration, the fidelity improves to $\frac{(F_0^{(d)})^{2^n}}{\sum_{r=0}^{d-1}(F_r^{(d)})^{2^n}}.$ Figure~\ref{FIG.6} shows the relationship between the improved fidelity and the number of iterations $n$ for the three-dimensional quantum system. In principle, the fidelity can be increased to unity through multiple rounds of iteration.

For general noise containing both qudit-flip and phase-flip errors, the noisy mixed state is given by $\tilde{\rho}_{AB}^{(d)}=\sum_{n,m=0}^{d-1}F_{n,m}^{(d)}|\Psi_{n,m}^{(d)}\rangle_{AB}\langle\Psi_{n,m}^{(d)}|$. After the qudit-flip purification step described in Sec. \ref{Sec3.1} with measurement results $|f_0\rangle_a|f_0\rangle_b$, $|f_1\rangle_a|f_{d-1}\rangle_b$, $\ldots$, $|f_{d-1}\rangle_a|f_1\rangle_b$, the fidelity of $|\Psi_{0,0}^{(d)}\rangle$ is $\frac{F_{0,0}^{(d)}\tilde{F}_0^{(d)}}{\sum_{n,m=0}^{d-1}F_{n,m}^{(d)}\tilde{F}_m^{(d)}}$. A subsequent phase-flip purification step in Sec. \ref{Sec3.2} further improves the fidelity to $\frac{F_{0,0}^{(d)}(\tilde{F}_0^{(d)})^2}{\sum_{n,m=0}^{d-1}F_{n,m}^{(d)}\tilde{F}_n^{(d)}\tilde{F}_m^{(d)}}$.  For the case where $F_{0,0}^{(d)}=\tilde{F}_0^{(d)}=F_0$, $F_{n,m}^{(d)}=\frac{1-F_0}{d^2-1}$ (except $F_{0,0}^{(d)})$, and $\tilde{F}_1^{(d)}=\tilde{F}_2^{(d)}=\cdots=\tilde{F}_{d-1}^{(d)}=\frac{1-F_0}{d-1}$,
the success condition remains
$F_{0,0}^{(d)}\in(\frac{1}{d},1]$ and the success probability is $\frac{1}{d}$.

In conclusion, we have proposed a scalable method for arbitrary high-dimensional EPPs for qudit-Bell states based on path superposition. By selecting appropriate single-qudit operations, our protocol can purify both qudit-flip errors and phase-flip errors. The proposed HDEPP requires fewer gate operations than existing ones \cite{miguel2018efficient}. Notably, higher-dimensional systems yield higher purification fidelity. Through iterative purification rounds, the fidelity can be improved to unity in principle.
Furthermore, we have developed EPPs for the frequency DOF using photonic hyperentanglement. This single-copy implementation circumvents the challenges associated with double-pair emission and multi-mode cases of photons, making it more experimentally feasible. The approach can also be  extended to other DOFs. Our work holds promise for applications in high-dimensional long-distance quantum communication, high-dimensional distributed quantum computing, and quantum networks.

\section*{Funding} \par

This work was supported by the National Natural Science Foundation of China under Grant No. 12505028 and Grant No. 62371038, and Science Research Project of Hebei Education Department under Grant No. QN2025054.

\section*{Disclosures} \par
The authors declare no conflicts of interest.

\section*{Data availability} \par

Data underlying the results presented in this paper are not publicly available at this time but maybe obtained from the authors upon reasonable request.

\end{document}